    \documentstyle[aps,preprint]{revtex}

\begin{document}


    \title{Duality equivalence between nonlinear self-dual and topologically 
    massive models\footnote{Dedicated to the memory of Prof. Juan Alberto Mignaco}}
    \author{ Anderson Ilha and Clovis Wotzasek }
    \address{Instituto de F\'{\i}sica\\ Universidade Federal do Rio de Janeiro\\ 
    Caixa Postal 68528, 21945 Rio de Janeiro, Rio de Janeiro, Brazil}
    \date{\today}
    \maketitle


    \begin{abstract}
    In this report we study the dual equivalence between the generalized 
    self-dual (SD) and topologically massive (TM) models.    To this end we linearize the model using an auxiliary field and apply a gauge embedding procedure to construct a gauge equivalent model. We clearly show that, under the above conditions, a nonlinear SD model always has a
duality equivalent TM 
    action.The general result obtained is then particularized 
    for a number of examples, including the Born-Infeld-Chern-Simons (BICS) model 
    recently discussed in the literature.
    \end{abstract}



    \section{INTRODUCTION}
    \label{sec:level1}

    In this report we are interested in the duality equivalence between models which
    are
    apparently different but nevertheless describe the same physical phenomenon, keeping
    invariant some properties such as the
    number of degrees of freedom, propagator and equations of motion.  The paradigm of
    this equivalence
    is the well known duality between the SD \cite{TPN} and MCS \cite{MCS} models in 2+1
    dimensions. This is 
    made possible by the introduction of the topological and gauge invariant Chern-Simons
    term 
    (CST) \cite{CS},  also responsible for essential features manifested by
    three-dimensional 
    field theories, such as parity breaking and anomalous spin \cite{??-01}. 

    The investigation of duality equivalence in three dimensions involving CST has had a
    long 
    and fruitful history, beginning when Deser and Jackiw used the master action concept
    to prove the dynamical equivalence between the SD  and MCS 
    theories \cite{DJ}, in this way proving the existence of a hidden symmetry in the SD
    version. This approach has been extensively used thereafter, providing an
    invaluable tool in the study of the planar physics phenomena and in the extension of
    the bosonization program from two to three dimensions with important
    phenomenological
    consequences \cite{varios}.

    Led by this well-known equivalence, we ask ourselves if these dualities can be
    extended in an arbitrary way, i.e., given a ``general'' nonlinear self dual model (NSD), what is its
    corresponding MCS-like dual
    equivalent? To answer this we use the auxiliary field technique to linearize the NSD model in terms of the $A^{2} = A_{\mu}A^{\mu}$ argument and employ an iterative embedding procedure \cite{NPB} to construct a
    gauge
    invariant theory out of the non-linear SD which leads to a general MCS model. As it is appropriate for
    a gauge embedding procedure, it produces changes in the nature of the constraints of the SD theory.
    However, instead of focusing on the constraints, we iteratively introduce counter-terms built with powers of the SD Euler
    vector \cite{PLB}.  Clearly, the resulting theory is on-shell equivalent with the original
    nonlinear SD model but, by construction, the result is gauge invariant.  Basically
    this involves disclosing, in the language of constraints, hidden gauge symmetries in
    such
    systems. The nonlinear SD can be considered as the gauge fixed version of the gauge
    theory.
    The latter reverts to the former under certain gauge fixing conditions, thus
    obtaining a deeper and more
    illuminating interpretation of these systems. The associated gauge theory is
    therefore to be
    considered as the ``gauge embedded'' version of the original second-class theory. The
    advantage in having a gauge theory lies in the fact that the underlying gauge
    symmetry allows
    us to establish a chain of equivalence among different models by choosing different
    gauge
    fixing conditions.

This paper is organized as follows. In the next section we start with a discussion of the linearization procedure that allows us to reduce the NSD model into an ordinary SD model plus a function of an auxiliary field. After that the dual transformation is performed and the final effective theory is finally obtained after the removal of the auxiliary field. Some examples are discussed in Section III and in Section IV we extend the present formalism to include coupling with fermionic matter. Our results are discussed in the last section.

    \section{Gauge Embedding}

    To derive our results we will consider the following nonlinear generalization of the Townsend-Pilch-Nieuwenhuizen
SD model,
    \begin{equation}
    {\cal L}_{NSD} = g(A^{2}) - \chi\,\frac{m}{2}\,
    \epsilon^{\mu\nu\lambda}\,A_{\mu}\,\partial_{\nu}
    A_{\lambda}\,,
    \label{AIS01}
    \end{equation} 
    where $m$ is a coupling constant playing the role of a mass parameter, $\chi$ is the chirality signal, assuming the values 
    $\chi = \pm 1$, and $g$ is a generic function of the model's basic field 
    $A_{\mu}$.  Note that $g$ depends explicity on 
    $A^{2} = A_{\mu}A^{\mu}$ only which, together with the existence of a linear representation (\ref{AIS03B}) below, are the only restrictions we will put on this function.

    It is useful to briefly clarify some properties exhibited by this model. The
    equations 
    of motion derived from Eq.(\ref{AIS01}) are given by
    \begin{equation}
    A_{\mu} = \frac{\chi}{2\,g^{\prime}}\,
    \epsilon_{\mu\nu\lambda}\,\partial^{\nu}A^{\lambda}\,,
    \label{AIS01B}
    \end{equation}
    where the prime denotes a derivative with respect to the $A^2$ argument.
    From
    these equations the following two relations can be verified,
    \begin{equation}
    \label{AIS01C}
    \partial_{\mu}A^{\mu} = \epsilon_{\mu\nu\lambda}\,
    \partial^{\mu}\left(\frac{\chi}{2\,g^{\prime}}\right)\,\partial^{\nu}A^{\lambda}\,,
    \end{equation}
    and
    \begin{equation}
    \label{AIS01D}
    \left(\Box + 4\,g^{\prime\,2}\right)\,A_{\mu} =
    \partial_{\mu}\left(\partial_{\nu}A^{\nu}\right) +
    g^{\prime}\,\partial^{\nu}\left(\frac{1}{g^{\prime}}\right)\,F_{\mu\nu}\,,
    \end{equation}
    where $F_{\mu\nu} = \partial_{\mu}A_{\nu} - \partial_{\nu}A_{\mu}$. Note that
    unless we have the linear SD model, $g(A^2) = \frac{m^2}{2}\,A^{2}$, the nonlinear
    SD model defined by Eq.(\ref{AIS01}) in general does not propagate a transverse
    massive mode. However, the nonlinear SD model possess a well-defined self-dual
    property in the same manner as in its linear counterpart. This can be seen as
    follows. Define a field dual to $A_{\mu}$ as
    \begin{equation}
    \label{AIS01E}
    {}^{\star}A_{\mu} \equiv \frac{1}{2\,g^{\prime}}\,\epsilon_{\mu\nu\lambda}\,
    \partial^{\nu}A^{\lambda}\, ,
    \end{equation}
    and repeat this dual operation to find that
    \begin{equation}
    \label{AIS01F}
    {}^{\star}\left({}^{\star}A_{\mu}\right) = 
    \frac{1}{2\,g^{\prime}}\,
    \epsilon_{\mu\nu\lambda}\,\partial^{\nu}\left(\frac{1}{2\,g^{\prime}}\,
    \epsilon^{\lambda\alpha\beta}\,\partial_{\alpha}A_{\beta}\right)\,,
    \end{equation}
    which can be rewritten as
    \begin{equation}
    {}^{\star}\left({}^{\star}A_{\mu}\right) =
    \frac{1}{4\,g^{\prime\,2}}\,
    \left[\partial_{\mu}\left(\partial_{\nu}A^{\nu}\right) + 
    2\,g^{\prime}\,\partial^{\nu}\left(\frac{1}{2\,g^{\prime}}\right)\,F_{\mu\nu}
    -\Box\,A_{\mu}\right]\,.
    \end{equation}
    Exploiting Eq.(\ref{AIS01D}) we have
    \begin{equation}
    \label{AIS01G}
    {}^{\star}\left({}^{\star}A_{\mu}\right) = A_{\mu}\,.
    \end{equation}
    thereby validating the
    definition of the dual field. Combining these results with Eq.(\ref{AIS01B}),
    we conclude that
    \begin{equation}
    \label{AIS01H}
    {}^{\star}A_{\mu} = \chi\,A_{\mu}\, ,
    \end{equation}
    hence, depending on the signature of $\chi$, the theory will correspond to a
    self-dual
    or an anti-self-dual model, irrespective of the particular form assumed by the 
    function $g(A^2)$.

    Next, let us deal with the nonlinear term. In order to take the nonlinearity of the NSD model (\ref{AIS01}) into account,
    within the gauge embedding procedure, we assume that the $g(A^2)$ term
    possess a linear representation, in terms of an ancillary field $\lambda$ (basically a Legendre transformation), such that
    \begin{equation}
    \label{AIS03B}
    g(A^2) \to \frac{A^2}{\lambda} + f(\lambda)\,,
    \end{equation}
    in the Lagrangian, with $f(\lambda)$ being an auxiliary function to be determined in the case by case basis. By writing the
    nonlinear
    SD model in this form we  have actually encapsulated all its former nonlinearity upon
    the field $\lambda$ and all we have now is a standard SD model, in terms of the basic
    field $A_{\mu}$,
    \begin{equation}
    {\cal L}_{\lambda} = \frac{A^2}{\lambda} + f(\lambda) - 
    \chi\,\frac{m}{2}\,\epsilon^{\mu\nu\lambda}\,A_{\mu}\,\partial_{\nu}A_{\lambda}\, ,
    \label{AIS02}
    \end{equation}
with a field dependent mass parameter.    The crucial point is, of course, how to find an appropriate $f(\lambda)$ for which
    Eq.(\ref{AIS03B}) holds true. To this end we find the variational solution $\bar\lambda$ of the Lagrangian (\ref{AIS03B}) for the auxiliary field 
    $\lambda$,
    \begin{equation}
    \label{AIS03}
    \left[f^{\prime}(\lambda) - \frac{A^2}{\lambda^2}\right]_{\lambda=\bar\lambda}=0 \, ,
    \end{equation}
that can be integrated as

 \begin{equation}
    f(\lambda) = \int^\lambda\,d\sigma\,\frac{1}{\sigma^{2}}\,
    A^2\left(\sigma\right)\,.
    \label{AIS033}
    \end{equation}
where we have relabeled $\bar\lambda\to\lambda$.

    The next step is to find the relation of the basic field $A_{\mu}$ with the auxiliary
    field $\lambda$ by an inverse Legendre transform. We then find, from Eq.(\ref{AIS03B})    \begin{equation}
    \left[g^{\prime}(A^2) - \frac{1}{\lambda}\right]_{A^2 =\bar{A^2}}=0\,,
    \label{AIS04}
    \end{equation}
and define, formally, a new function 
    $h(A^2) \equiv g^{\prime}(A^{2})$ such that its inverse produces the desired relation upon use of
Eq.(\ref{AIS04}),
    \begin{equation}
    A^2(\lambda) = h^{-1}\left(\frac{1}{\lambda}\right)\,.
    \label{AIS05}
    \end{equation}
    Bringing this result in  Eq.(\ref{AIS033}), we have
    \begin{equation}
    f(\lambda) = \int^\lambda\,d\sigma\,\frac{1}{\sigma^{2}}\,
    h^{-1}\left(\frac{1}{\sigma}\right)\,,
    \label{AIS06}
    \end{equation}
    less an integration constant which is of no consequence for the equations of motion. 

    Once the linear representation is found, we may return to the discussion of duality equivalence.
Turning back to the Eq.(\ref{AIS02}), our solution
    then may follow from the iterative embedding procedure \cite{NPB,PLB} or any other approach such as the master action of Deser and Jackiw \cite{DJ}. Since we are interested also in the coupling with dynamical matter, here we follow Refs.\cite{NPB,PLB} where the basic idea is to modify the original SD model (\ref{AIS01}) with counter-terms
    built with powers of the Euler vector of the SD model which automatically guarantees the on-shell equivalence. Besides, we look for the special form of the counter-terms that allows one to
    lift a global symmetry of the SD model into its local form as,

    \begin{equation}
    A_\mu \to A_\mu + \partial_\mu \epsilon \, ,
    \label{AIS065}
    \end{equation}
with the lift of the global parameter $\epsilon$.
    To this end we compute the Euler vector

    \begin{equation}
    K^{\mu} = \frac{2}{\lambda}\,A_{\mu} - 
    \chi\,m\,\epsilon_{\mu\nu\lambda}\,\partial^{\nu}A_{\lambda}\, ,
    \label{AIS07}
    \end{equation}
and treat it as a global Noether charge, bringing it back into the SD action, with the help of an auxiliary gauge
    field $B_{\mu}$ and define a first-iterated Lagrangian as
    \begin{equation}
    {\cal L}^{(1)} = {\cal L}_{\lambda} - B_{\mu}\,K^{\mu}\,.
    \label{AIS08}
    \end{equation} 
    To cancel the variation of the original action ${\cal L}_{\lambda}$ it is convenient to choose the
    transformation property of the auxiliary field $B_{\mu}$ such
    that 
    \begin{equation}
    \delta\,B_{\mu} = \delta\,A_{\mu} = \partial_\mu\epsilon\,.
    \label{AIS09}
    \end{equation}
    Under this transformation the action (\ref{AIS08}) changes as
    \begin{equation}
    \delta\,{\cal L}^{(1)} = -\,\delta\,\left(\frac{1}{\lambda}\,B_{\mu}\,B^{\mu}\right)
    + 
    m\,B_{\mu}\,\epsilon^{\mu\nu\lambda}\,\partial\,\delta\,A_{\lambda}\,.
    \label{AIS10}
    \end{equation}
    Under the vector gauge transformations Eq.(\ref{AIS09}) the second term in the 
    r.h.s. of Eq.(\ref{AIS10}) vanishes identically leading to a second iterated
    Lagrangian
    \begin{equation}
    {\cal L}^{(2)} = {\cal L}^{(1)} + \frac{1}{\lambda}\,B_{\mu}\,B^{\mu}\,,
    \label{AIS11}
    \end{equation}
    that is gauge invariant under the combined action of $A_\mu$ and $B_\mu$, Eq.(\ref{AIS09}). We have therefore succeeded in 
    transforming the global SD theory (\ref{AIS02}) into a locally invariant gauge 
    theory. We may now take advantage of the Gaussian character of the auxiliary 
    field $B_{\mu}$ to rewrite Eq.(\ref{AIS11}) as an effective action depending only 
    on the fields $A_{\mu}$ and $\lambda$,
    \begin{equation}
    {\cal L}_{eff} = {\cal L}_{\lambda} - \frac{\lambda}{4}\,K_{\mu}\,K^{\mu}\,.
    \label{AIS12}
    \end{equation}
    It is straightforward to see, using the structures of the Euler vector
    Eq.(\ref{AIS07}), that this effective model corresponds to
    \begin{equation}
    {\cal L}_{eff} = \chi\,\frac{m}{2}\,\epsilon^{\mu\nu\lambda}\,A_{\mu}\,\partial_{\nu}
    A_{\lambda} - \frac{m^{2}}{8}\,\lambda\,F_{\mu\nu}\,F^{\mu\nu} + f(\lambda)\,,
    \label{AIS13}
    \end{equation}
which is clearly gauge invariant.
    After solving for the auxiliary field $\lambda$, we will restore the nonlinearity 
    inherent in the model, getting a functional form $H(F_{\mu\nu}F^{\mu\nu})$
    dependent solely on the quantity $F_{\mu\nu}F^{\mu\nu}$, 
    \begin{equation}
    {\cal L}_{TM} = \chi\,\frac{m}{2}\,\epsilon^{\mu\nu\lambda}\,A_{\mu}\,\partial_{\nu}
    A_{\lambda} - H\left(F_{\mu\nu}\,F^{\mu\nu}\right)\,,
    \label{AIS13B}
    \end{equation}
    which is the general topologically massive theory dual to the nonlinear SD model 
    Eq.(\ref{AIS01}). Note how $H\left(F_{\mu\nu}\,F^{\mu\nu}\right)$ is directly 
    related to $g(A_{\mu}A^{\mu})$ through the auxiliary function Eq.(\ref{AIS06}).
    In fact, in the examples which we will discuss below we will find that the same functional form is
    present in both SD-TM models, which can provide powerful insights in more
    complicated cases.

    Although we have started from a known nonlinear SD model and then determined its dual
    TM counterpart, we could have done it backwards as well.  Starting from a known
    TM model, given by Eq.(\ref{AIS13B}) we must find the corresponding auxiliary
    function $f(\lambda)$ such that
    \begin{equation}
    \label{AIS13C}
    f(\lambda) - \frac{m^{2}}{8}\,\lambda\,F_{\mu\nu}\,F^{\mu\nu} = 
    H\left(F_{\mu\nu}\,F^{\mu\nu}\right)\,,
    \end{equation}
    holds true. The function found is thus used in Eqs.(\ref{AIS03}) and (\ref{AIS04}),
    so that $g(A^2)$ can be  determined. This procedure would then be alternative to the gauge-fixing.

    \section{Examples}

    In this section we illustrate the procedures outlined above by giving some 
    examples which show the power and generality of the method by investigating 
    the dual correspondence between some nonlinear SD and TM models. These
    include a rational-power generalization of the usual SD model, the recently 
    discussed Born-Infeld-Chern-Simons (BICS) model and a logarithmic SD model. Although we will
    start from the gauge non invariant model towards the gauge invariant one, i.e,
    by unfixing the gauge freedom as we proceed,  as mentioned above, 
    this could also be done backwards, starting from the gauge-invariant TM model 
    and breaking the gauge freedom towards the gauge non-invariant SD model.

    \subsection{The rational self dual model}

    Consider the following self-dual model, given by
    \begin{equation}
    {\cal L}_{NSD} =
    \frac{q\,\beta^2}{p}\,\left(\frac{1}{\beta^2}\,A_{\mu}\,A^{\mu}\right)^{p/q} - 
    \chi\;\frac{m}{2}\,\epsilon^{\mu\nu\lambda}\,A_{\mu}\,\partial_{\nu}A_{\lambda}\,,
    \label{AIS22}
    \end{equation}
    where $p,q \in Z$ but $p/q \not= \{1,1/2\}$. Here the constant $\beta$ was inserted for dimensional reasons. 
    When $p=q$ this of course reduces to the 
    usual SD model while $q=2p$ is a troublesome one and will be discussed separately. Using the definitions from the preceding section, we have

    \begin{equation}
    g(A^{2}) = \frac{q\,\beta^2}{p}\,\left(\frac{1}{\beta^2}\,A_{\mu}\,A^{\mu}\right)^{p/q} = 
    \frac{1}{\lambda}\, \left(A_{\mu}\,A^{\mu}\right) + f(\lambda)\,.
    \label{AIS23}
    \end{equation}
    From this expression we can relate the basic field $A_{\mu}$ with the 
    auxiliary field $\lambda$ through Eq.(\ref{AIS04}),

    \begin{equation}
    A_{\mu}\,A^{\mu} = \beta^2\,\left(\frac{1}{\lambda}\right)^{\frac q{p-q}}\,,
    \label{AIS24}
    \end{equation}
    and use this relation in Eq.(\ref{AIS033}) to find the expression for the 
    auxiliary function $f(\lambda)$,

    \begin{equation}
    f(\lambda) = 
    -\beta^2\,\left(\frac{q-p}{p}\right)\,
    \left(\lambda\right)^{\frac p{q-p}}\,.
    \label{AIS25}
    \end{equation}
    We can now use this expression to write down the effective model. It is given by
    \begin{equation}
    {\cal L}_{eff} = \chi\,\frac{m}{2}\,\epsilon^{\mu\nu\lambda}\,A_{\mu}\,
    \partial_{\nu}A_{\lambda} - \frac{m^{2}}{8}\,\lambda\,F_{\mu\nu}\,F^{\mu\nu} 
    -\beta^2\,\left(\frac{q-p}{p}\right)\,
    \left(\lambda\right)^{\frac p{q-p}}\,.
    \label{AIS25B}
    \end{equation}
    Solving for the auxiliary field $\lambda$, we will have finally
    \begin{eqnarray}
    {\cal L}_{TM} &=& \chi\, \frac{m}{2}\,\epsilon^{\mu\nu\lambda}\,A_{\mu}\,
    \partial_{\nu}A_{\lambda} - \beta^{\frac{p-q}{2p-q}}\,\left(\frac{2p - q}{q}\right)\,
    \left(\frac{m^2}{8}\,F_{\mu\nu}F^{\mu\nu}\right)^{\frac p{2p -q}}\nonumber\\
&=& \chi\, \frac{m}{2}\,\epsilon^{\mu\nu\lambda}\,A_{\mu}\,
    \partial_{\nu}A_{\lambda} - \beta^{\frac{s-r}{s}}\,\left(\frac{s}{2r-s}\right)\,
    \left(\frac{m^2}{8}\,F_{\mu\nu}F^{\mu\nu}\right)^{\frac r{s}}\,,
    \label{AIS28}
    \end{eqnarray}
    which is the TM theory dual to the NSD model Eq.(\ref{AIS22}) after the relabeling $p\to r$ and $q\to 2r-s$. Notice that the rational SD model is mapped under duality to a rational TM model. It is clear that the case $p=q$ gives us back the usual SD-MCS duality, as it should but the case $q=2p$ becomes ill defined in (\ref{AIS28}). It is valid though both in (\ref{AIS25}) and (\ref{AIS25B}) where the auxiliary function $f(\lambda$) becomes linear in $\lambda$. We discuss this case below. However, it is interesting to note 
    that the rational SD model Eq.(\ref{AIS22}), when $p=Nq$ for large value of $N$ gives rise to a square-rooted TM model.
Similarly, if we let $p\to r$ and $q\to 2r-s$ such that $r= Ns$ then, for large values of $N$ we obtain a square-rooted SD model. Also interesting to observe is that self-duality in the sense that the ratios before and after dualization are the same $p/q = r/s$ are only satisfied if $p=q$ or $p=0$, i.e., the SD model or the pure Chern-Simons model.

Let us consider the case $q=2p$ explicitly.  The NSD model is,

 \begin{equation}
    {\cal L}_{NSD} =
    2\,\beta^2\,\left(\frac{1}{\beta^2}\,A_{\mu}\,A^{\mu}\right)^{1/2} - 
    \chi\;\frac{m}{2}\,\epsilon^{\mu\nu\lambda}\,A_{\mu}\,\partial_{\nu}A_{\lambda}\,,
    \label{AIS222}
    \end{equation}
while the effective dual equivalent theory is,

 \begin{equation}
    {\cal L}_{eff} = \chi\,\frac{m}{2}\,\epsilon^{\mu\nu\lambda}\,A_{\mu}\,
    \partial_{\nu}A_{\lambda} - \lambda\,\left[\frac{m^{2}}{8}\,F_{\mu\nu}\,F^{\mu\nu} 
    - \beta^2\right]\,.
    \label{AIS223}
    \end{equation}
We see that in this case the auxiliary function $\lambda$ becomes a Lagrange multiplier 
imposing the condition $F^2\sim const.$ as a constraint. In fact, this constraint is
also present in Eq.(\ref{AIS222}). To see this, let us use the equations of motion of both
Eqs.(\ref{AIS222}) and (\ref{AIS223}),
\begin{equation}
A_{\mu} = \frac{1}{2}\,\chi\,m\,\sqrt{\frac{1}{\beta^2}\,A_{\sigma}A^{\sigma}}\,
\epsilon_{\mu\nu\lambda}\,F^{\nu\lambda}\,,
\label{AIS224}
\end{equation}
and
\begin{equation}
\chi\,\epsilon_{\mu\nu\lambda}\,\partial^{\nu}A^{\lambda} + 
\frac{m}{2}\,\lambda\,\partial^{\sigma}F_{\sigma\mu} = 0\,,
\label{AIS225}
\end{equation}
respectively. Multiplying both sides of Eq.(\ref{AIS224}) by  
$\epsilon_{\mu\alpha\beta}F^{\alpha\beta}$, we have
\begin{equation}
\chi\,m\,\sqrt{\frac{1}{\beta^2}\,A_{\sigma}A^{\sigma}}\,F_{\mu\nu}F^{\mu\nu} = 
2\,\epsilon^{\mu\alpha\beta}\,
A_{\mu}\,F_{\alpha\beta}\,.
\label{AIS226}
\end{equation}
On the other hand, by Eq.(\ref{AIS224}) we have also
\begin{equation}
A_{\mu}\,A^{\mu} = \frac{1}{4}\,\chi\,m\,
\sqrt{\frac{1}{\beta^2}\,A_{\sigma}A^{\sigma}}\,
\epsilon^{\mu\alpha\beta}\,A_{\mu}\,F_{\alpha\beta}\,.
\label{AIS227}
\end{equation}
Using the above equation in Eq.(\ref{AIS226}) we have finally
\begin{equation}
\frac{m^2}{8}\,F_{\mu\nu}F^{\mu\nu} - \beta^2 = 0\,,
\label{AIS228}
\end{equation}
which is the constraint implemented in Eq.(\ref{AIS223}). We can also explicitly show
the equivalence between the equations of motion Eqs.(\ref{AIS224}) and (\ref{AIS225}).
Note that we can rewrite the later as
\begin{equation}
\epsilon^{\mu\nu\sigma}\,\partial_{\nu}\left[
\chi\,A_{\sigma} - \lambda\,\frac{m}{2}\,\epsilon_{\sigma\alpha\beta}\,
\partial^{\alpha}A^{\beta}\right] = 0\,.
\label{AIS229}
\end{equation}
Since this expression is gauge invariant, it implies that
\begin{equation}
A_{\sigma} - \lambda\,\chi\, \frac{m}{2}\,\epsilon_{\sigma\alpha\beta}\,
\partial^{\alpha}A^{\beta} = \partial_{\sigma}\Omega\,,
\label{AIS230}
\end{equation}
where $\Omega$ is an arbitrary function. Now we can fix the gauge in such a manner that
$\partial_{\sigma}\Omega = 0$. Moreover, using the relation between $\lambda$ and $A_{\mu}$,
Eq.(\ref{AIS24}), we get
\begin{equation}
A_{\sigma} - \chi\,\frac{m}{2}\,
\sqrt{\frac{1}{\beta^2}\,A_{\rho}A^{\rho}}\,
\epsilon_{\sigma\alpha\beta}\, \partial^{\alpha}A^{\beta} = 0\,,
\label{AIS231}
\end{equation}
which is exactly  Eq.(\ref{AIS224}). In other words, the selfdual equation of motion
is a gauge fixed, canonically equivalent to the topologically massive equation 
Eq.(\ref{AIS225}).

    \subsection{the logarithmic SD model}

    Another interesting possibility is given by the logarithmic SD model. Here we consider the following action
    \begin{equation}
    {\cal L}_{NSD} = \beta^2\,\ln\left(\frac{1}{\beta^2}\,A_{\mu}\,A^{\mu}\right) - 
    \chi\,\frac{m}{2}\,\epsilon^{\mu\nu\lambda}\,A_{\mu}\,
    \partial_{\nu}A_{\lambda}\,,
    \label{AIS29}
    \end{equation}
    where $\beta$ is a parameter inserted for dimensional reasons. This model has a linear representation as

    \begin{equation}
  \beta^2\,\ln\left(\frac{1}{\beta^2}\,A_{\mu}\,A^{\mu}\right)
    = \frac{1}{\lambda}\, \left(A_{\mu}\,A^{\mu}\right) + f(\lambda)\,.
    \label{AIS30}
    \end{equation}
    with

    \begin{equation}
    f(\lambda) = \beta^2\,\ln(\lambda)\,. 
    \label{AIS32}
    \end{equation}
    The effective theory resulting from this procedure is given by
    \begin{equation}
    {\cal L}_{eff} = \chi\,\frac{m}{2}\,\epsilon^{\mu\nu\lambda}\,A_{\mu}\,
    \partial_{\nu}A_{\lambda} - \lambda\,\frac{m^2}{8}\,F_{\mu\nu}\,F^{\mu\nu} + 
    \beta^{2}\,\ln(\lambda)\,.
    \label{AIS33}
    \end{equation}
    Solving this model for the auxiliary field $\lambda$, we have
    \begin{equation}
    {\cal L}_{TM} = \chi\,\frac{m}{2}\,\epsilon^{\mu\nu\lambda}\,A_{\mu}\,
    \partial_{\nu}A_{\lambda} - 
    \beta^2\,\ln\left[\frac{1}{\beta^2}\,\left(\frac{m^2}{8}\,F_{\mu\nu}\,
    F^{\mu\nu}\right)\right]\,,
    \label{AIS34}
    \end{equation}
    less a constant that can safely be set to zero since it will give no dynamical
    contribution.
    This is the TM model dual to the NSD model Eq.(\ref{AIS29}).  Notice in particular the logarithmic dependence of the TM model, exactly the same as the NSD 
    model.

    \subsection{the BICS model}

    As our final example let us study the dual correspondence between the nonlinear SD model proposed in \cite{khare} and BICS 
    model. This relationship was found in an indirect way by means of 
    Hamiltonian techniques. Here we employ the gauge embedding procedure, starting
    by the following model
    \begin{equation}
    \label{AIS35}
    {\cal L}_{NSD} = \beta^{2}\,
    \sqrt{1 + \frac{1}{m^2\beta^2}\,\left(A_{\mu}\,A^{\mu}\right)} - \chi\,
    \frac{m}{2}\,\epsilon^{\mu\nu\lambda}\,A_{\mu}\,\partial_{\nu}A_{\lambda}\,.
    \end{equation}
    As before,  $\beta$ is a parameter inserted for dimensional reasons. Note that
    the above model has in the limit $\beta \rightarrow \infty$ the usual SD
    model.  Therefore, using
    our notation we have
    \begin{equation}
    g(A^2) = \beta^{2}\,
    \sqrt{1 + \frac{1}{m^2\beta^2}\,\left(A_{\mu}\,A^{\mu}\right)}
    = \frac{1}{\lambda}\, \left(A_{\mu}\,A^{\mu}\right) + f(\lambda)\,,
    \label{AIS36}
    \end{equation}
    which, using Eq.(\ref{AIS04}) gives
    \begin{equation}
    A_{\mu}\,A^{\mu} = \beta^{2}\,\left(
    \frac{\lambda^2}{4\,m^2} - m^2\right)\,.
    \label{AIS37}
    \end{equation}
    This identity can be used to evaluate the auxiliary function $f(\lambda)$ by
    means of Eq.(\ref{AIS03}),
    \begin{equation}
    f(\lambda) = \beta^{2}\,\left(\frac{\lambda}{4\,m^2} +
    \frac{m^2}{\lambda}\right)\,,
    \label{AIS38}
    \end{equation}
    The effective model is thus given by
    \begin{equation}
    {\cal L}_{eff} = \chi\,\frac{m}{2}\,\epsilon^{\mu\nu\lambda}\,
    A_{\mu}\,\partial_{\nu}A_{\lambda} - \lambda\,\left[
    \frac{m^2}{8}\,F_{\mu\nu}\,F^{\mu\nu} - \frac{\beta^2}{4\,m^2}\right]
    + \frac{\beta^2\,m^2}{\lambda}\,,
    \label{AIS39}
    \end{equation}
    which can be solved for the auxiliary field $\lambda$ to produce the BICS model
    \begin{equation}
    {\cal L}_{TM} = \beta^{2}\,
    \sqrt{1-\frac{1}{2\beta^{2}}\,F_{\mu\nu}\,F^{\mu\nu}} + \chi\,
    \frac{m}{2}\,\epsilon^{\mu\nu\lambda}\,A_{\mu}\,\partial_{\nu}A_{\lambda}\,.
    \label{AIS40}
    \end{equation}
Notice that in the limit $\beta \rightarrow \infty$ we recover the usual TM model. Therefore, as expected, the duality relationship is also respected in this limit.
It is interesting to observe, once again, the same functional relation for the general SD model and the general TM model.


    \section{Coupling with dynamical matter}

    In this section we apply the iterative procedure to construct a gauge invariant
    theory out of the NSD model
    coupled to dynamical matter fields, generalizing the treatment proposed in \cite{USP}. As in the free case, to guarantee
    equivalence with the starting 
    non-invariant theory, we only use counter-terms vanishing in the space of solutions of
    the model. To be specific let us consider the minimal coupling of the NSD vector field to dynamical fermions, so that the Lagrangian becomes
    \begin{equation}
    {\cal L}^{(0)} = {\cal L}_{NSD} - e\,A_{\mu}\,J^{\mu} + {\cal L}_{D}\,,
    \label{AIS41}
    \end{equation}
    where $J_{\mu} = \bar{\psi}\,\gamma_{\mu}\,\psi$, and the superscript index 
    is the iterative counter. Here the Dirac Lagrangian is 
    \begin{equation}
    {\cal L}_{D} = \bar{\psi}\,\left(i\,\slash\!\!\!\partial - M\right)\,\psi\,,
    \label{AIS42}
    \end{equation}
    where $M$ is the fermion mass.  To implement the Noether embedding we follow 
    the usual track and compute the Euler vector for the Lagrangian 
    ${\cal L}_{NSD}$ given by Eq.(\ref{AIS29}), showing the presence of the 
    fermionic current,
    \begin{equation}
    K^{\mu} = \frac{2}{\lambda}\,A^{\mu} - m\,
    \epsilon^{\mu\nu\lambda}\,\partial_{\nu}A_{\lambda} - e\,J^{\mu}\,.
    \label{AIS43}
    \end{equation}
    The effective theory that comes out after the dualization procedure is 
    implemented as before, except with the Euler vector replaced by 
    Eq.(\ref{AIS43}), yielding
    \begin{eqnarray}
    {\cal L}_{eff} = f(\lambda) &+&
    \chi\,\frac{m}{2}\,\epsilon^{\mu\nu\lambda}\,A_{\mu}\,
    \partial_{\nu}A_{\lambda}  + {\cal L}_{D}  \nonumber \\
    &-&\frac{\lambda}{2}\,\left[
    \frac{m^2}{4}\,F_{\mu\nu}\,F^{\mu\nu} + \frac{e^2}{2}\,J_{\mu}\,J^{\mu}
    + \chi\,e\,m\,\epsilon^{\mu\nu\lambda}\,J_{\mu}\,\partial_{\nu}A_{\lambda}
    \right]\,.
    \label{AIS44}
    \end{eqnarray}
    A simple inspection shows that the minimal coupling of the nonlinear SD model was
    replaced by a 
    nonminimal magnetic Pauli type coupling and the appearance of a Thirring like current-current term,
    which  
    are characteristic features of Chern-Simons dualities involving matter couplings.

    The auxiliary function $f(\lambda)$ can be computed by using 
    Eqs.(\ref{AIS03}) and (\ref{AIS04}) as before.
    By solving the equations of motion for the auxiliary field $\lambda$, we
    will get the topologically massive model
    \begin{equation}
    {\cal L}_{TM} = \chi\,\frac{m}{2}\,\epsilon^{\mu\nu\lambda}\,A_{\mu}\,
    \partial_{\nu}A_{\lambda}  + H(F_{\mu\nu},J_{\mu}) +
    {\cal L}_{D}\,,
    \label{AIS46}
    \end{equation}
    where $H(F_{\mu\nu},J_{\mu})$ is a functional form depending on the field strength
    $F_{\mu\nu}$
    and the current couplings. 

As an example consider the rational SD model treated above. Using (\ref{AIS25}) in (\ref{AIS44}) above gives

\begin{equation}
    {\cal L}_{TM} ={\cal L}_{D} +
\chi\, \frac{m}{2}\,\epsilon^{\mu\nu\lambda}\,A_{\mu}\,
    \partial_{\nu}A_{\lambda} - \beta^{\frac{s-r}{s}}\,\left(\frac{s}{2r-s}\right)\,
    \left[
    \frac{m^2}{8}\,F_{\mu\nu}\,F^{\mu\nu} + \frac{e^2}{4}\,J_{\mu}\,J^{\mu}
    + \frac{\chi\,e\,m}{2}\,\epsilon^{\mu\nu\lambda}\,J_{\mu}\,\partial_{\nu}A_{\lambda}
    \right]^{\frac r{s}}\, .\nonumber
\end{equation}
All the other examples follow straightforwardly.

    \section{conclusions}

    In this paper we studied the dual equivalence between the nonlinear generalization of the self dual and the
    topologically massive models in 2+1 
    dimensions, in the context of the Noether embedding procedure, which provides a clear
    physical meaning of the
    duality equivalence. This is accomplished by linearizing the nonlinear terms by means
    of a auxiliary field,
    which can be eliminated later on in order to restore the full nonlinearity of the NSD and the generalized MCS
    models.  
    The usual SD-MCS dual equivalence are naturally contained in this results, including the couplings with dynamical matter. Some
    examples are discussed that both
    clarify and prove the power of the gauge embedding technique to deal with duality equivalence.

\noindent ACKNOWLEDGMENTS: This work is partially supported by CNPq, CAPES, FAPERJ and
FUJB, Brazilian Research Agencies.

    \end{document}